\documentclass[aps, prd, floats, floatfix, twocolumn, superscriptaddress, nofootinbib, showpacs, letterpaper, 10pt]{revtex4-1}

\usepackage{graphicx}
\usepackage{amsmath}
\usepackage{amsfonts}
\usepackage{amssymb}
\usepackage{amstext}
\usepackage{tensor}
\usepackage[english]{babel}
\usepackage{tabularx}
\usepackage{array}
\usepackage{helvet}
\usepackage{microtype}
\usepackage[pdftex,
  pdftitle={The Shape Dependence of Vainshtein Screening},
  pdfauthor={Jolyon K. Bloomfield and Clare Burrage},
  bookmarks,bookmarksopen=false,
  pdfstartview={FitH},
  linktocpage=true
]{hyperref}

\def\arraystretch{2.4}

\begin{document}

\title{The Shape Dependence of Vainshtein Screening}

\author{Jolyon K. Bloomfield}
\affiliation{Center for Theoretical Physics, Laboratory for Nuclear Science, and Department of Physics, Massachusetts Institute of Technology, Cambridge, MA 02139, USA}
\email{jolyon@mit.edu}

\author{Clare Burrage}
\affiliation{School of Physics and Astronomy, University of Nottingham, Nottingham, NG7 2RD, UK}
\email{clare.burrage@nottingham.ac.uk}

\author{Anne-Christine Davis}
\affiliation{Department of Applied Mathematics and Theoretical Physics, Centre for Mathematical Sciences, Cambridge, CB3 0WA, UK}
\email{acd@damtp.cam.ac.uk}

\date{\today}

\begin{abstract}
Scalar field theories that possess a Vainshtein mechanism are able to dynamically suppress the associated fifth forces in the presence of massive sources through derivative non-linearities. The resulting equations of motion for the scalar are highly non-linear and therefore very few analytic solutions are known. Here we present a brief investigation of the structure of Vainshtein screening in symmetrical configurations, focusing in particular on the spherical, cylindrical and planar solutions that are relevant for observations of the cosmic web. We consider Vainshtein screening in both the Galileon model, where the non-linear terms involve second derivatives of the scalar, and a k-essence theory, where the non-linear terms involve only first derivatives of the scalar. We find that screening, and consequently the suppression of the scalar force, is most efficient around spherical sources, weaker around cylindrical sources and can be absent altogether around planar sources.
\end{abstract}


\maketitle

\section{Introduction}

Are there new, light degrees of freedom associated with the physics explaining the current acceleration of the expansion of the universe?  The simplest explanation of the observed current behaviour of the universe is the introduction of a cosmological constant, however the required value of this constant continues to defy explanation in a quantum theory. Alternative theories almost universally introduce new light scalars \cite{Copeland:2006wr,Clifton:2011jh} that would mediate long range fifth forces, and yet no such force has been seen to date. In the absence of an explanation for why such scalars would be forbidden from interacting with matter fields, scalar fields are required to possess a screening mechanism in order to dynamically hide the resulting force from observations.  Screening mechanisms rely on the presence of non-trivial self interactions of the scalar field in order to change the behaviour of the field dynamically on differing scales and in differing environments. We can classify screening mechanisms depending on the type of self-interactions that lead to the screening: $\phi$ screening, which includes chameleon \cite{Khoury:2003aq}, symmetron \cite{Hinterbichler:2010es,Olive:2007aj,Pietroni:2005pv} and varying dilaton mechanisms \cite{Brax:2010gi,Brax:2011ja}; $\partial \phi$ screening, which includes k-essence \cite{Brax:2012jr}, k-mouflage \cite{Babichev:2009ee} and D-BIonic screening mechanisms \cite{Burrage:2014uwa}; and $\partial \partial \phi$ screening, which is a property of the Galileon and generalised Galileon models \cite{Nicolis2008,Kimura:2011dc,Narikawa:2013pjr,Kase:2013uja}.

In the two latter cases, the screening of the scalar field around a source occurs when the field gradients become large and the derivative non-linearities begin to dominate the evolution of the scalar field.  This is known as Vainshtein screening \cite{Vainshtein:1972sx}, and the distance scale within which the screening occurs is known as the Vainshtein radius. In this work, we discuss both types of Vainshtein screening, considering theories that rely on non-linearities in both first and second derivatives of the scalar field. In each case, we work with a specific model to illustrate the screening behaviour: we use the D-BIonic scalar, an example of $\partial \phi$ screening, and the Galileon model, an example of $\partial\partial \phi$ screening. These are particularly interesting examples of screening, as in the limit where the coupling to matter vanishes, both theories possess symmetries which protect the self-interactions of the scalar field from quantum corrections \cite{Luty:2003vm,Nicolis:2004qq,Burrage:2014uwa}; introducing the coupling to matter only mildly breaks this symmetry. While this property makes these theories particularly interesting to study, the phenomenology of each screening mechanism is common to the broader class of theories.

Screening mechanisms require non-linear interactions, and therefore the scalar field profile can be sensitive to the shape of a source in a way that Newtonian gravitational forces are not. For all of the screening mechanisms mentioned above, the phenomenon of screening has been demonstrated for static, spherically symmetric sources. To a good approximation, this configuration describes many objects in our universe, including galaxy halos, stars and planets. The efficiency of screening in such conditions is invoked to evade fifth-force constraints in the vicinity of such objects.

However, given the apparent necessity of introducing light degrees of freedom for cosmological purposes, we would like to consider environments in which screening is not so efficient, in order to place tighter constraints on scalar field theories with screening. For this reason it is important to study screening beyond the static, spherically symmetric approximation. Previous work in this field has investigated screening behaviors about spherical bodies in two-body systems and in slowly-rotating regimes in a fully relativistic description \cite{Hiramatsu:2012xj,Chagoya:2014fza}. In this work we study the presence (or absence) of Vainshtein screening for a number of static, one-dimensional systems with completely different geometries. Vainshtein screening is a particularly interesting target for this investigation as it is already known that no screening of the Galileon field occurs around a planar source \cite{Brax:2011sv}\footnote{Asymptotic solutions for the chameleon field profile around an ellipsoidal source are also known \cite{2012PhRvL.108v1101J}.}. We leave the possibility of relaxing the static assumption for future work.

We treat Galileon and D-BIonic theories in Sections \ref{sec:galileon} and \ref{sec:dbionic} respectively. In the Galileon case, we review known results in spherical and planar symmetry, and present new solutions in cylindrical symmetry. In the D-BIonic case, we present new solutions in planar and cylindrical symmetry, and review known results in spherical symmetry. In Section \ref{sec:discussion} we discuss the implications of these results and their connections with cosmological observations.

\section{The Galileon} \label{sec:galileon}

The flat space Galileon action introduced by Nicolis \textit{et al.} \cite{Nicolis2008} is given by
\begin{align}\label{eq:fullaction}
  S = \int d^4x \sqrt{-g} & \bigg[ - \frac{1}{2} {\cal L}_2 - \frac{1}{2 \Lambda^3} {\cal L}_3 - \frac{\lambda_4}{2\Lambda^6} {\cal L}_4
\nonumber\\
  & \qquad - \frac{\lambda_5}{2\Lambda^9} {\cal L}_5 + \frac{\beta \phi}{M_P} \tensor{T}{^\mu_\mu}\bigg]
\end{align}
where
\begin{subequations}
\begin{align}
  {\cal L}_2 &= (\nabla \phi)^2
\\
  {\cal L}_3 &= \square \phi (\nabla \phi)^2
\\
  {\cal L}_4 &= (\nabla \phi)^2 \left[ (\square \phi)^2 - \nabla_\mu \nabla_\nu \phi \nabla^\mu \nabla^\nu \phi \right]
\\
  {\cal L}_5 &= (\nabla \phi)^2 \big[ (\square \phi)^3 - 3 (\square \phi) \nabla_\mu \nabla_\nu \phi \nabla^\mu \nabla^\nu \phi
\nonumber\\
  & \qquad \qquad \quad + 2 \nabla^\mu \nabla_\nu \phi \nabla^\nu \nabla_\rho \phi \nabla^\rho \nabla_\mu \phi \big]
\end{align}
\end{subequations}
with $(\nabla \phi)^2 = \nabla_\mu \phi \nabla^\mu \phi$ and $\square \phi = \nabla^\mu \nabla_\mu \phi$, and where $M_P = 1/\sqrt{8 \pi G}$ is the reduced Planck mass. The first four terms in this action are invariant under the Galileon symmetry
\begin{align}
\phi\rightarrow \phi +b_{\mu}x^{\mu}+c
\end{align}
for arbitrary constants $b_{\mu}$ and $c$, up to total derivative terms.  Although a tadpole term is also compatible with the symmetry, we do not include it here.
The covariant form was first given by Deffayet \textit{et al.} \cite{Deffayet2009a}, however in this work we restrict our attention to situations where the curvature is weak. For the static, non-relativistic sources that we investigate, corrections due to spacetime curvature will be governed by the size of the Newtonian potential and its derivatives. We thus expect the theory described by the action \eqref{eq:fullaction} to be sufficient for our purposes. Furthermore, we will use flat metrics to investigate solutions to the scalar field equations. We expect that corrections to these solutions due to spacetime curvature effects will go as $O(\Phi)$, which for our purposes are negligible.

The final term in the action \eqref{eq:fullaction} represents a conformal coupling to the trace of the stress-energy tensor of the matter sector, which breaks the Galileon symmetry.  The presence of this coupling means that test particles of mass $m$ experience a Galileon force of the form
\begin{align}
\vec{F}_{\phi}= - m \frac{\beta}{M_P}\vec{\nabla} \phi\,.
\end{align}

Neglecting the coupling to matter, the Galileon action can be alternatively expressed as an action in $D$ dimensions in the following manner, as described by Deffayet \textit{et al.} \cite{Deffayet2009b}.
\begin{align} \label{eq:fancyaction}
  S = \int d^D x \sqrt{-g} \phi \; \sum_{n=1}^D \lambda_n \tensor*[^n]{A}{^{\mu_1 \ldots \mu_n}_{\nu_1 \ldots \nu_n}} \Pi_{i=1}^n \nabla_{\mu_i} \nabla^{\nu_i} \phi
\end{align}
Here, we absorb various coefficients into the coupling constants $\lambda_n$. The tensor $\tensor*[^n]{A}{^{\mu_1 \ldots \mu_n}_{\nu_1 \ldots \nu_n}}$ is defined as the contraction of $D-n$ indices between two epsilon tensors as
\begin{align}
  \tensor*[^n]{A}{^{\mu_1 \ldots \mu_n}_{\nu_1 \ldots \nu_n}} = \epsilon^{\mu_1 \ldots \mu_n \alpha_{n+1} \ldots \alpha_D} \epsilon_{\nu_1 \ldots \nu_n \alpha_{n+1} \ldots \alpha_D}
\end{align}
where
\begin{align}
  \epsilon^{\mu_1 \ldots \mu_n} = - \frac{1}{\sqrt{-g}} \delta_1^{[\mu_1} \delta_2^{\mu_2} \ldots \delta_n^{\mu_n]} \,.
\end{align}
Note that $\tensor*[^n]{A}{}$ is completely antisymmetric on the top indices and the bottom indices.

Different $n$ correspond to different order Galileon terms. The $n=1$ term is the quadratic Galileon, the $n=2$ term the cubic Galileon, and so on. In this form, it is obvious that there are a finite number of Galileon terms, as the $\tensor*[^n]{A}{}$ tensor can only antisymmetrize over a number of indices equal to the spacetime dimension, and no more. In particular, in four-dimensional spacetime, the highest order Galileon possible is the quintic Galileon.

\subsection{Static Solutions}

We begin by looking at the Galileon equation of motion in Cartesian coordinates. Starting from the Galileon part of the  action \eqref{eq:fancyaction}, the equation of motion can be expressed as
\begin{align}
  \sum_{n=1}^D \lambda_n \tensor*[^n]{A}{^{\mu_1 \ldots \mu_n}_{\nu_1 \ldots \nu_n}} \Pi_{i=1}^n \partial_{\mu_i} \partial^{\nu_i} \phi = 0 \,.
\end{align}
In this form, it is straightforward to see that for a given order $n$, the terms in the equation of motion will be zero if the number of Cartesian coordinates that $\phi$ depends on is less than $n$ (modulo terms of the form $b_\mu x^\mu$, which vanish when twice differentiated). For example, if $\phi = \phi(x)$, then only the $n=1$ term will survive, as for $n>1$, all terms contain products of partial derivatives of $\phi$ that differentiate with respect to $y$, $z$ or $t$ and therefore vanish.

This suggests that for static configurations in planar symmetry, we expect only the quadratic Galileons to contribute. In cylindrical symmetry, the quadratic and cubic terms contribute, as $\phi(r) = \phi(\sqrt{x^2 + y^2})$ depends on both $x$ and $y$. Spherical symmetry will receive contributions from the quadratic, cubic and quartic Galileon terms. The quintic term can never contribute to static solutions; only configurations that depend non-trivially on $x$, $y$, $z$ and $t$ are influenced by the quintic term. Alternatively, notice that when flat dimensions are present in the metric and the Galileon configuration is independent of this dimension, the configuration is also a solution of a theory with fewer dimensions, where correspondingly fewer nontrivial Galileon terms exist in the action. In all static solutions, the flat time dimension could well have been integrated out of the action (effectively removing the quintic term), and similarly for further symmetric solutions. This greatly simplifies the structure of the equations of motion when appropriate symmetries are present. It also allows for the possibility of breaking the degeneracy between the Galileon parameters $\lambda_i$ by studying configurations with different spatial symmetries. This argument is more generally true for the class of theories which possess $(\partial\partial\phi)$ screening, known as generalised Galileons, because terms which include second derivatives of $\phi$ always have to enter with the same index structure as the Galileon terms in order to avoid the presence of ghost degrees of freedom.

Let us now look towards solving the full equations of motion under the assumption of static configurations. Screening is most important in non-relativistic scenarios where all of our searches for deviations from Newtonian gravity are carried out, including laboratory searches for fifth forces, and solar systems constraints on deviations from the $r^{-2}$ force law. These are the tests that screening mechanisms are designed to avoid. In this regime, the mass energy completely dominates the stress-energy tensor, and pressure and anisotropic stresses are negligible. We thus assume a matter configuration consisting only of an energy density with stress-energy tensor $\tensor{T}{^\mu_\mu} = \tensor{T}{^0_0} = -\rho$. The equation of motion from the action \eqref{eq:fullaction} is written in covariant notation as the following, where we neglect the quintic terms which vanish under the static assumption.
\begin{align}
  \frac{\beta}{M_P} \rho &= \square \phi + \frac{1}{\Lambda^3} \left[(\square \phi)^2 - (\nabla_\mu \nabla_\nu \phi) (\nabla^\mu \nabla^\nu \phi)\right]
\nonumber \\
   & \qquad + \frac{\lambda_4}{\Lambda^6} \big[ (\square \phi)^3
   - 3 \square \phi (\nabla_\mu \nabla_\nu \phi) (\nabla^\mu \nabla^\nu \phi)
\nonumber \\
   & \qquad \qquad \quad
   + 2 (\nabla^\mu \nabla_\nu \phi) (\nabla^\nu \nabla_\gamma \phi) (\nabla^\gamma \nabla_\mu \phi)\big]
\end{align}
We use step-functions for our energy density profiles, as we are primarily concerned with the exterior field solutions for the scalar field (such as outside a planet/star). We show below that the exterior solutions only ever depend on the total enclosed mass (or appropriate mass density in cylindrical or planar symmetries), which further justifies restricting our investigation to sources of constant density.

Far away from a source we expect the field to be close to the vacuum solution $\phi \approx \mathrm{const}$.  Therefore gradients of the field will be small and the non-linear terms in the equation of motion can be neglected when compared with $\square \phi$.  If Vainshtein screening occurs then as we approach a source, gradients of the field will increase and the non-linear terms will begin to dominate, changing the form of the scalar field profile.  The distance scale within which the non-linear terms dominate is known as the Vainshtein radius.

\subsubsection{Planar Symmetry}
We begin by investigating planar symmetry using the metric
\begin{align}
  ds^2 = - dt^2 + dx^2 + dy^2 + dz^2 \,.
\end{align}
We choose $\phi = \phi(z)$, and assume that $\rho = \rho(z)$ also. Such a scenario was first considered for the Galileon in \cite{Brax:2011sv}. Only the quadratic and coupling terms survive in the equation of motion.
\begin{align}
  \frac{\beta}{M_P} \rho(z) = \partial_z^2 \phi
\end{align}
For concreteness, let $\rho(z) = \rho_0$ between $\pm z_0$ and zero outside, and choose the zero of the potential to be $\phi(0)=0$. We can then integrate to obtain
\begin{align}
  \partial_z \phi &= \left\{
  \begin{array}{cc}
  \displaystyle \frac{\beta \rho_0}{M_P} z \qquad & |z| < z_0
\\
  \displaystyle \frac{\beta \rho_0}{M_P} z_0 \qquad & |z| \ge z_0
  \end{array}
  \right.
\end{align}
and
\begin{align}
  \phi &= \left\{
  \begin{array}{lc}
  \displaystyle \frac{\beta \rho_0}{2 M_P} z^2 \qquad & |z| < z_0
\\
  \displaystyle \frac{\beta \rho_0 z_0}{M_P} \left(z - \frac{z_0}{2}\right) \qquad & |z| \ge z_0
  \end{array}
  \right.
\end{align}
where $\partial_z \phi = 0$ at the origin by symmetry. The absence of the scale $\Lambda$ from these expressions clearly indicates that no non-linear or screening effects are present. As the gravitational force outside the plane has magnitude $F_G = 2 \rho_0 z_0 m/M_P^2$, the ratio of the scalar force to the corresponding gravitational force $F_\phi / F_G$ is given by
\begin{align}
  \frac{F_\phi}{F_G} = 2 \beta^2 \,.
\end{align}

\subsubsection{Cylindrical Symmetry}
We next investigate cylindrical symmetry, using the metric
\begin{align}
  ds^2 = -dt^2 + dr^2 + r^2 d\theta^2 + dz^2 \,.
\end{align}
We take $\phi = \phi(r)$ as well as $\rho = \rho(r)$. The quadratic, cubic and coupling terms contribute to the equation of motion
\begin{align}
  \frac{\beta}{M_P} \rho(r) = \phi'' + \frac{\phi'}{r} + \frac{2 \phi' \phi''}{r \Lambda^3}
\end{align}
where we use primes to denote derivatives with respect to $r$.

Let us consider a cylinder with constant mass density $\rho = \rho_0$ for $r < r_0$, and zero outside. The equation of motion can be rearranged into
\begin{align}
  \frac{\beta}{M_P} r \rho(r) = (r \phi')' + \frac{(\phi'^2)'}{\Lambda^3}
	\label{eq:cylallgal}
\end{align}
which can be straightforwardly integrated over $r$. We choose our boundary conditions to be $\phi(0)=0$, and cylindrical symmetry demands $\phi^{\prime}(0)=0$.

If  the cubic term is absent ($\Lambda\rightarrow \infty$), we have
\begin{align}
  \phi' =
  \left\{
  \begin{array}{lc}
  \displaystyle \frac{\beta \rho_0 r}{2 M_P} \qquad & r < r_0
\\
  \displaystyle \frac{\beta \rho_0 r_0^2}{2 M_P r} \qquad & r \ge r_0
  \end{array}
  \right.
\end{align}
giving the expected $\sim1/r$ force law in the exterior of the source. The gravitational force sourced by the same cylindrical object is $F_G = m \rho_0 r_0^2 / 4 M_P^2 r$, again yielding the ratio
\begin{align}
  \frac{F_\phi}{F_G} = 2 \beta^2 \,.
\end{align}
The corresponding scalar potential is
\begin{align}
  \phi =
  \left\{
  \begin{array}{lc}
  \displaystyle \frac{\beta \rho_0 r^2}{4 M_P} \qquad & r < r_0
\\
  \displaystyle \frac{\beta \rho_0 r_0^2}{4 M_P} \left[1 + 2 \ln\left(\frac{r}{r_0}\right) \right] \qquad & r \ge r_0
  \end{array}
  \right. \,.
\end{align}

We now turn to the full equation of motion. Solving Eq. \eqref{eq:cylallgal} for $\phi'$ yields
%
\begin{align}
  \phi' =
  \left\{
  \begin{array}{lc}
  \displaystyle \frac{\Lambda^3 r}{2} \left(\sqrt{1 + \frac{r_v^2}{r_0^2}} - 1\right) \qquad & r < r_0
  \\
  \displaystyle \frac{\Lambda^3 r}{2} \left(\sqrt{1 + \frac{r_v^2}{r^2}} - 1\right) & r \ge r_0
  \end{array}
  \right.
\end{align}
where the Vainshtein radius, within which the non-linear terms dominate the behaviour of the scalar, is
\begin{align}
  r_v = \sqrt{\frac{2 \beta \rho_0 r_0^2}{M_P \Lambda^3}} = \sqrt{\frac{2 \beta \lambda}{\pi M_P \Lambda^3}}
\end{align}
where $\lambda = \pi r_0^2 \rho_0$ is the linear mass density.  We have chosen a positive sign outside the square root to ensure that we recover the $1/r$ unscreened force law at large distances from the source.  We also impose continuity of $\phi^{\prime}$ at $r=r_0$.

Integrating one last time, we obtain the scalar potentials
\begin{align}
  \phi &= \frac{\Lambda^3}{4} \left(\sqrt{1 + \frac{r_v^2}{r_0^2}} - 1\right) r^2
\end{align}
for $r < r_0$ and
\begin{align}
  \phi &= \frac{\Lambda^3}{4} \bigg[r^2 \left(\sqrt{1 + \frac{r_v^2}{r^2}} - 1\right)
\nonumber\\
  & \qquad + r_v^2 \ln \left(\frac{r + \sqrt{r^2 + r_v^2}}{r_0 + \sqrt{r_0^2 + r_v^2}}\right)
  \bigg]
\end{align}
for $r \ge r_0$\footnote{The logarithm can also be written as a pair of arcsinh functions as
\begin{align*}
  \ln \left(\frac{r + \sqrt{r^2 + r_v^2}}{r_0 + \sqrt{r_0^2 + r_v^2}}\right) = \mathrm{arcsinh} \left(\frac{r}{r_v}\right) - \mathrm{arcsinh} \left(\frac{r_0}{r_v}\right) \,.
\end{align*}}.

Deep inside the Vainshtein radius $r_0 < r \ll r_v$, the scalar force saturates at a constant magnitude $F_{\phi} = m \beta \Lambda^3 r_v /2 M_P$, meaning that in this region the scalar force is suppressed compared to the gravitational force sourced by the same cylindrical object by
\begin{align}
  \frac{F_\phi}{F_G} = 4 \beta^2 \frac{r}{r_v} \,.
\end{align}

The behaviour of the screening around a cylindrical source is illustrated in Fig. \ref{fig:plot}.

\subsubsection{Spherical Symmetry}
Finally we turn to spherical symmetry, where we use the metric
\begin{align}
  ds^2 = -dt^2 + dr^2 + r^2 d\theta^2 + r^2 \sin^2\theta d\phi^2
\end{align}
and take $\phi = \phi(r)$ and $\rho = \rho(r)$. Spherically symmetric solutions for the Galileon were first studied in \cite{Nicolis2008,Burrage:2010rs}. Galileon terms up to quartic order contribute to the equation of motion
\begin{align}
  \frac{\beta}{M_P} \rho(r) = \phi'' + \frac{2 \phi'}{r} + \frac{2 \phi'^2}{r^2 \Lambda^3} + \frac{4 \phi' \phi''}{r \Lambda^3} + \frac{6 \lambda_4 \phi'^2 \phi''}{r^2 \Lambda^6}
\end{align}
where a prime now indicates differentiation with respect to the radial coordinate of the spherically symmetric metric.

We begin by rearranging the equation of motion into the following form.
\begin{align}
  \frac{\beta}{M_P} r^2 \rho(r) = (r^2 \phi')' + \frac{2 (r \phi'^2)'}{\Lambda^3} + \frac{2 \lambda_4 (\phi'^3)'}{\Lambda^6}
\end{align}
Let us take $\rho = \rho_0$ for $r < r_0$, and again choose $\phi(0) = 0$ as the zero of our potential. Spherical symmetry yields $\phi'(0) = 0$. Generally speaking, this equation is intractable, and full analytic solutions are only known for particular values of $\lambda_4$. However in all cases it is possible to determine the asymptotic form of the solutions.

When the cubic and quartic terms are turned off ($\Lambda \rightarrow \infty$), the field derivatives are simply given by
\begin{align}
  \phi' =
  \left\{
  \begin{array}{lc}
  \displaystyle \frac{\beta \rho_0}{3 M_P} r \qquad & r < r_0
  \\
  \displaystyle \frac{\beta \rho_0 r_0}{3 M_P} \frac{r_0^2}{r^2} & r \ge r_0
  \end{array}
  \right.
\end{align}
which can be integrated to give
\begin{align}
  \phi =
  \left\{
  \begin{array}{lc}
  \displaystyle \frac{\beta M}{8 \pi M_P r_0} \frac{r^2}{r_0^2} \qquad & r < r_0
  \\
  \displaystyle \frac{\beta M}{4 \pi M_P r_0} \left(\frac{3}{2} - \frac{r_0}{r}\right) & r \ge r_0
  \end{array}
  \right.
\end{align}
where we let $M = 4 \pi r_0^3 \rho_0 / 3$. As expected, this exhibits a $1/r^2$ force that is disallowed by solar system constraints unless $\beta \ll 1$. The magnitude of the gravitational force for $r>r_0$ is $F_G = Mm / 8 \pi M_P^2 r^2$, again giving the ratio
\begin{align}
  \frac{F_\phi}{F_G} = 2 \beta^2 \,.
\end{align}

When the cubic term is present but the quartic term vanishes ($\lambda_4 \rightarrow 0$), $\phi'$ becomes
%
%
\begin{align}
  \phi' =
  \left\{
  \begin{array}{lc}
  \displaystyle \frac{\Lambda^3}{4} r \left(\sqrt{1 + \frac{r_v^3}{r_0^3}} - 1\right) \qquad & r < r_0
  \\
  \displaystyle \frac{\Lambda^3}{4} r \left(\sqrt{1 + \frac{r_v^3}{r^3}} - 1\right) & r \ge r_0
  \end{array}
  \right.
\end{align}
where we have identified the Vainshtein radius as
\begin{align} \label{eq:cubicrv}
  r_v = \left(\frac{8 \beta \rho_0 r_0^3}{3 M_P \Lambda^3}\right)^{1/3} = \left(\frac{2 \beta M}{\pi M_P \Lambda^3}\right)^{1/3} \,.
\end{align}
 Deep inside the Vainshtein radius, the scalar force goes as $\sim 1/\sqrt{r}$, with the ratio of the Galileon force to the corresponding gravitational force being
\begin{align}
  \frac{F_\phi}{F_G} = 4 \beta^2 \left( \frac{r}{r_v}\right)^{3/2} \,.
\end{align}
The expression for $\phi'$ can be integrated to obtain
\begin{align}
  \phi = \frac{\Lambda^3}{8} r^2 \left(\sqrt{1 + \frac{r_v^3}{r_0^3}} - 1\right)
\end{align}
for $r < r_0$, and
\begin{align}
  \phi &= \frac{\Lambda^3}{8} \bigg(
  r^2 \left[\sqrt{1 + \frac{r_v^3}{r^3}} - 1\right]
\nonumber\\
  & \qquad \qquad + 3 \sqrt{r_v^3 r_0} \bigg[\sqrt{\frac{r}{r_0}} \;\; \tensor[_2]{F}{_1}\left(\frac{1}{6}, \frac{1}{2}; \frac{7}{6}; - \frac{r^3}{r_v^3}\right)
\nonumber\\
  & \qquad \qquad \qquad \qquad - \tensor[_2]{F}{_1}\left(\frac{1}{6}, \frac{1}{2}; \frac{7}{6}; - \frac{r_0^3}{r_v^3}\right) \bigg]
  \bigg)
\end{align}
for $r \ge r_0$, where $\tensor[_2]{F}{_1}(a,b;c;d)$ is the hypergeometric function.

The presence of the quartic term requires solving the following cubic polynomial equation.
\begin{align} \label{eq:generalspherical}
  r^2 \phi' + \frac{2 r \phi'^2}{\Lambda^3} + \frac{2 \lambda_4 \phi'^3}{\Lambda^6} =
  \left\{
  \begin{array}{lc}
  \displaystyle \frac{\beta M}{4 \pi M_P} \frac{r^3}{r_0^3} \qquad & r < r_0
  \\
  \displaystyle \frac{\beta M}{4 \pi M_P} & r \ge r_0
  \end{array}
  \right.
\end{align}
In general, these solutions are unpleasant. However, the distance scale controlling when the quartic Galileon term becomes important in the equation of motion can still be identified.

Due to stability constraints \cite{Nicolis2008}, the coefficients appearing in the action are limited to $\Lambda>0$ and $0 \le \lambda_4 \le \frac{2}{3}$. For $\lambda_4>0$ but within these limitations, there will be a region about the origin in which the quartic term dominates, followed by a region in which the cubic term dominates, and subsequently a region in which the quadratic term dominates \cite{Burrage:2010rs}. The crossover at which the cubic and quadratic terms are equally important is just the cubic Vainshtein radius \eqref{eq:cubicrv}.

At the crossover radius $r_{v4}$ when the cubic and quartic terms are equally important, we have
\begin{align}
  \phi' = \frac{r_{v4} \Lambda^3}{\lambda_4} \,.
\end{align}
Substituting this back in the equation of motion to solve for $r_{v4}$, we obtain
\begin{align}
  r_{v4} = \left(\frac{\lambda_4^2}{32}\right)^{1/3} \left(\frac{2 \beta M}{\pi M_P \Lambda^3}\right)^{1/3}
\end{align}
where we neglect the subdominant quadratic term. The quantity to the right here is just the cubic Vainshtein radius \eqref{eq:cubicrv}.
Deep inside this Vainshtein radius $r_0 < r \ll r_{v4}$, $\phi'$ saturates at the constant value
\begin{align}
  \phi' = \frac{2^{1/3} \Lambda^3}{\lambda_4} r_{v4}
\end{align}
and the scalar force is suppressed compared to the corresponding gravitational force by
\begin{align}
  \frac{F_\phi}{F_G} = \beta^2 \lambda_4 2^{-2/3} \frac{r^2}{r_{v4}^2} \,.
\end{align}

\begin{figure*}[t]
  \begin{center}
    \includegraphics{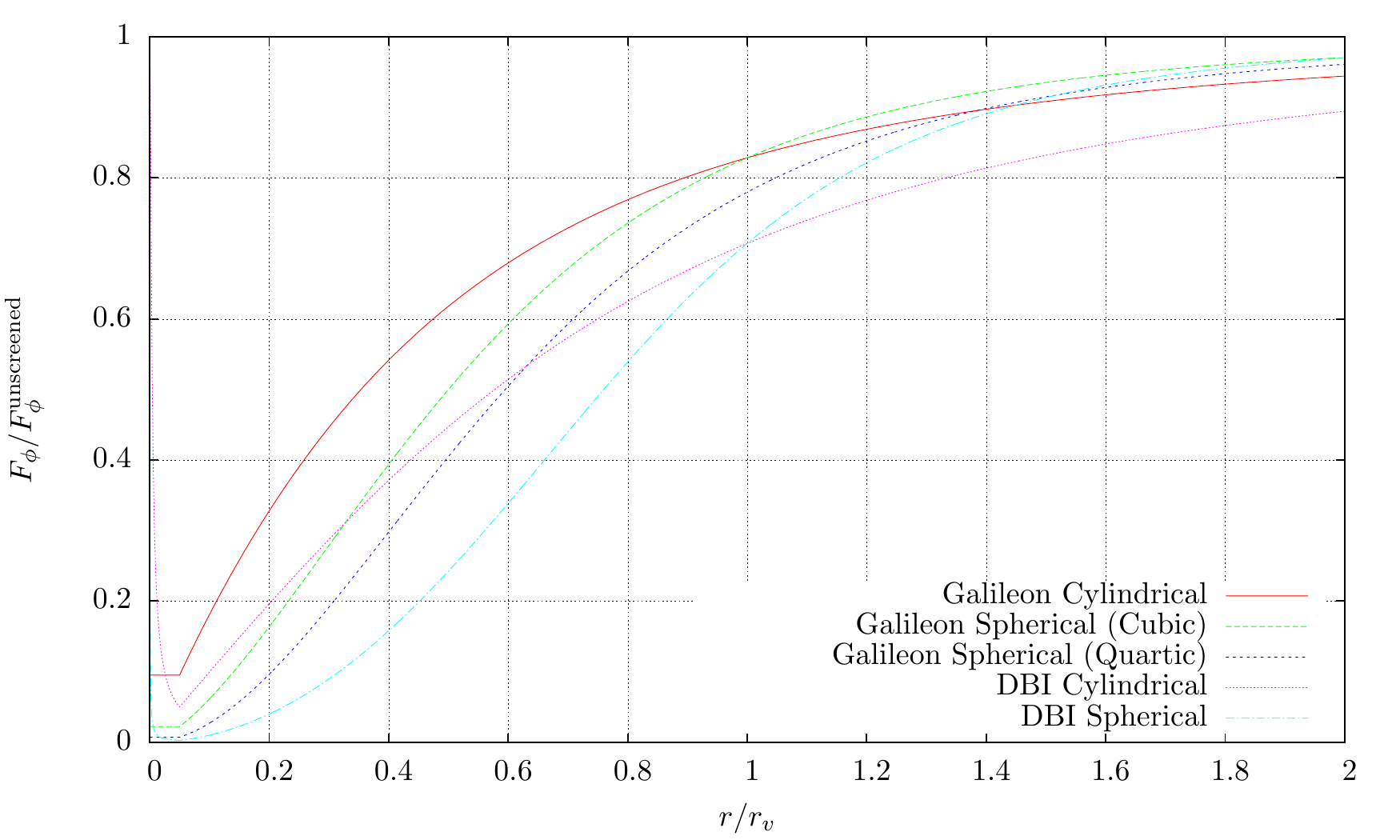}
    \caption{Plot of the screening fraction for cylindrical and spherical solutions, including both Galileon and D-BIonic models. The screening fraction is the ratio of the screened force to the unscreened (quadratic only) solution. In the case of the D-BIonic models, we used the quadratic Galileon as the unscreened reference (corresponding to the leading order term). To convert to the ratio $F_\phi / F_G$, simply multiply by $2 \beta^2$. For the purpose of this plot, $r_0 = 0.05 r_v$ was used. For the quartic Galileon case, we show the analytic result for $\lambda_4 = 2/3$. The behavior of the DBI case near the origin simply reflects the interior behavior of the fields and is not of significant interest (the ratio approaches unity at $r=0$).}
    \label{fig:plot}
  \end{center}
\end{figure*}

A particularly nice analytic solution in the quartic case can be found for $\lambda_4 = 2/3$.
\begin{align}
  \phi' =
  \left\{
  \begin{array}{lc}
  \displaystyle
  \frac{\Lambda^3}{2} r \left[\left(1 + \frac{r_v^3}{r_0^3} \right)^{1/3} - 1 \right]
  \qquad & r < r_0
  \\
  \displaystyle
  \frac{\Lambda^3}{2} r \left[\left(1 + \frac{r_v^3}{r^3} \right)^{1/3} - 1 \right]
  & r \ge r_0
  \end{array}
  \right.
\end{align}
The Vainshtein radius here is
\begin{align}
  r_v = \left(\frac{3}{4}\right)^{1/3} \left(\frac{2 \beta M}{\pi M_P \Lambda^3}\right)^{1/3}
\end{align}
which is approximately 91\% the size of the case for the cubic term alone. Note that in this limiting case, there is only one screened regime rather than the two described above; this arises because the quadratic, cubic and quartic terms are all equally important at this radius. In this case, deep inside the Vainshtein radius, the force saturates at
\begin{align}
  F_\phi = \frac{m \beta \Lambda^3 r_v}{2 M_P}
\end{align}
which yields a scalar to gravitational force ratio of
\begin{align}
  \frac{F_\phi}{F_G} = 6 \beta^2 \frac{r^2}{r_v^2} \,.
\end{align}
This solution is included as the quartic case in Fig. \ref{fig:plot}.

\section{The D-BIon} \label{sec:dbionic}

We now look at the behavior of a model that exhibits $\partial \phi$ screening. The D-BIonic model \cite{Burrage:2014uwa} has the following action.
\begin{align}
  S = \int d^4x \sqrt{-g} \left[ \Lambda^4 \sqrt{1 - \frac{(\nabla \phi)^2}{\Lambda^4}} + \frac{\beta \phi}{M_P} \tensor{T}{^\mu_\mu} \right]
	\label{eq:DBIaction}
\end{align}
Compared to the standard DBI form, both the overall sign of the first term in this action and the sign of $(\nabla \phi)^2$ have been flipped.  This is necessary to achieve screening, and it is straightforward to check that when the square root is expanded the scalar kinetic term has the correct sign for the theory to be free of ghosts. The DBI form means that the first term in the action is invariant under the following transformation of the field and the coordinates.
\begin{align}
  \tilde{\phi}(\tilde{x}) &= \gamma(\phi(x)+\Lambda^2 v_{\mu}x^{\mu})\;,
\\
  \tilde{x}^{\mu} &= x^{\mu} +\frac{\gamma-1}{v^2}v^{\mu}v_{\nu}x^{\nu}+\gamma v^{\mu}\frac{\phi(x)}{\Lambda^2}
\end{align}

The leading order term in Eq. \eqref{eq:DBIaction} expanded around $(\nabla \phi)^2 = 0$ is equivalent to the quadratic Galileon term by itself, so around any matter distribution we expect the same asymptotic behavior for the field profile as in the Galileon situation; in particular, we expect an attractive scalar force. The coupling term is identical to the Galileon coupling, and so the relationship between the scalar force and the gradient of the scalar is also identical.

The equation of motion resulting from the action \eqref{eq:DBIaction} is simply
\begin{align}
  \nabla_\mu \left(\frac{\nabla^\mu \phi}{\sqrt{1 - (\nabla \phi)^2 / \Lambda^4}} \right) = - \frac{\beta}{M_P} \tensor{T}{^\mu_\mu} \,.
\end{align}
We now investigate the static symmetric solutions as we did for the Galileons.

\subsection{Static Solutions}

As previously, we investigate situations with stress-energy tensor $\tensor{T}{^\mu_\mu} = -\rho$.

\subsubsection{Planar Symmetry}

Assuming that $\rho = \rho(z)$ and $\phi = \phi(z)$, the equation of motion becomes
\begin{align}
  \partial_z \left( \frac{\partial_z \phi}{\sqrt{1 - (\partial_z \phi)^2 / \Lambda^4}} \right) = \frac{\beta \rho}{M_P} \,.
\end{align}
Again, we take $\rho(z) = \rho_0$ between $\pm z_0$ and zero outside, and choose the zero of the potential to be $\phi(0)=0$. We can then integrate to obtain
\begin{align}
  \partial_z \phi &= \left\{
  \begin{array}{lc}
  \displaystyle
  \pm \frac{\Lambda^2}{\sqrt{1 + z_\ast^2/z^2}}
  \qquad & |z| < z_0
\\
  \displaystyle
  \pm \frac{\Lambda^2}{\sqrt{1 + z_\ast^2/z_0^2}}
  \qquad & |z| \ge z_0
  \end{array}
  \right.
\end{align}
where $z_\ast = M_P \Lambda^2 / \beta \rho_0$ is a characteristic length scale. Here, we take the positive (negative) root for $z > 0$ ($z < 0$) to obtain the appropriate asymptotics, and to ensure the continuity of $\partial_z \phi$. These expressions can be integrated to obtain the following.
\begin{align}
  \phi &= \left\{
  \begin{array}{lc}
  \displaystyle
  \Lambda^2 z_\ast \left(\sqrt{1 + \frac{z^2}{z_\ast^2}} - 1\right)
  \qquad & |z| < z_0
\\
  \displaystyle
  \Lambda^2 \left(
  \frac{z z_0 + z_\ast^2}{\sqrt{z_0^2 + z_\ast^2}}
  - z_\ast
  \right)
  \qquad & |z| \ge z_0
  \end{array}
  \right.
\end{align}
As is the case for the Galileon (and also purely canonical scalar fields), the scalar force is independent of $z$. However, unlike the Galileon, the strength of the force is not purely fixed by the coupling strength $\beta$. If $z_{\ast} \gg z_0$ then the D-BIon non-linearities are always subdominant, but if the density and size of the planar source are such that $z_{\ast} \ll z_0$, then the force is smaller than it would be in a theory with no non-linearities. The scale $z_{\ast}$ can still be thought of as the Vainshtein distance scale for this system. However, because the force around a planar source is constant with distance, we find that sources are either always screened if the width of the source is smaller than the Vainshtein scale $z_{\ast}$, or always unscreened if the width is larger than the Vainshtein scale.

\subsubsection{Cylindrical Symmetry}
We take $\phi = \phi(r)$ as well as $\rho = \rho(r)$. The equation of motion becomes
\begin{align}
  \partial_r \left( \frac{r \phi'}{\sqrt{1 - \phi'^2 / \Lambda^4}} \right) = \frac{\beta r \rho}{M_P}
\end{align}
where we use primes to denote derivatives with respect to the cylindrical radial coordinate $r$.

Let us again consider a cylinder with constant mass density $\rho = \rho_0$ for $r < r_0$. The equation of motion can be integrated over $r$ and solved for $\phi'$ to obtain the following.
\begin{align}
  \phi' &= \left\{
  \begin{array}{lc}
  \displaystyle
  \pm \frac{\Lambda^2}{\sqrt{1 + r_0^4 / r^2 r_v^2}}
  \qquad & r < r_0
\\
  \displaystyle
  \pm \frac{\Lambda^2}{\sqrt{1 + r^2 / r_v^2}}
  \qquad & r \ge r_0
  \end{array}
  \right.
\end{align}
Here, the Vainshtein radius is
\begin{align}
  r_v = \frac{\lambda_0 \beta}{2 \pi M_P \Lambda^2}
\end{align}
where $\lambda_0 = \pi r_0^2 \rho_0$ is the linear mass density. Again, we choose the positive roots by matching to the appropriate asymptotic form, and requiring continuity at $r_0$. We can integrate to obtain $\phi(r)$.
\begin{align}
  \phi &= \left\{
  \begin{array}{l}
  \displaystyle
  \frac{\Lambda^2 r_0^2}{r_v}\left(\sqrt{1 + \frac{r^2 r_v^2}{r_0^4}} - 1\right)
  \qquad \qquad r < r_0
\\
  \displaystyle
  \frac{\Lambda^2 r_0^2}{r_v} \left(\sqrt{1 + \frac{r_v^2}{r_0^2}} - 1
  + \frac{r_v^2}{r_0^2} \ln \left[\frac{r + \sqrt{r^2 + r_v^2}}{r_0 + \sqrt{r_0^2 + r_v^2}}\right]
  \right)
  \end{array}
  \right.
\end{align}
Here, the second expression is for $r > r_0$. This expression, particularly outside the object, bears a striking resemblance to the corresponding Galileon expression.

Deep inside the Vainshtein radius ($r_0 < r \ll r_v$), the scalar force saturates at $F_\phi = - m \beta \Lambda^2 / M_P$, giving a scalar to gravitational force ratio of
\begin{align}
  \frac{F_\phi}{F_G} = 2 \beta^2 \frac{r}{r_v}
\end{align}
which is the same as the Galileon case up to a factor of two.

\subsubsection{Spherical Symmetry}
We take $\phi = \phi(r)$ as well as $\rho = \rho(r)$, where $r$ is now the spherical radius. The equation of motion becomes
\begin{align}
  \partial_r \left( \frac{r^2 \phi'}{\sqrt{1 - \phi'^2 / \Lambda^4}} \right) = \frac{\beta r^2 \rho}{M_P}
\end{align}
where we use primes to denote derivatives with respect to $r$.

We again consider a sphere with constant mass density $\rho = \rho_0$ for $r < r_0$. The equation of motion can be integrated over $r$ and solved for $\phi'$ to obtain the following.
\begin{align}
  \phi' &= \left\{
  \begin{array}{lc}
  \displaystyle
  \pm \frac{\Lambda^2}{\sqrt{1 + r_0^6 / r^2 r_v^4}}
  \qquad & r < r_0
\\
  \displaystyle
  \pm \frac{\Lambda^2}{\sqrt{1 + r^4 / r_v^4}}
  \qquad & r \ge r_0
  \end{array}
  \right.
\end{align}
Here, the Vainshtein radius is
\begin{align}
  r_v = \sqrt{\frac{\beta M}{4 \pi M_P \Lambda^2}}
\end{align}
where $M = 4 \pi r_0^3 \rho_0 / 3$. Again, we choose the positive roots by matching to the appropriate asymptotic form, and applying continuity at $r_0$. We can integrate to obtain $\phi(r)$. For $r < r_0$, we have
\begin{align}
  \phi &= \frac{\Lambda^2 r_0^3}{r_v^2}\left(\sqrt{1 + \frac{r^2 r_v^4}{r_0^6}} - 1\right)
\end{align}
while for $r > r_0$, the integral again yields hypergeometric functions.
\begin{align}
  \phi = {}&\frac{\Lambda^2 r_0^3}{r_v^2}\left(\sqrt{1 + \frac{r_v^4}{r_0^4}} - 1\right)
\nonumber\\
  & - \frac{\Lambda^2 r_v^2}{r_0} \bigg[ \frac{r_0}{r} \; \tensor[_2]{F}{_1}\left(\frac{1}{4}, \frac{1}{2}; \frac{5}{4}; - \frac{r_v^4}{r^4}\right)
\nonumber\\
  & \qquad \qquad
  - \tensor[_2]{F}{_1}\left(\frac{1}{4}, \frac{1}{2}; \frac{5}{4}; - \frac{r_v^4}{r_0^4}\right)
  \bigg]
\end{align}
Again, this bears a striking resemblance to the solution for the cubic Galileon in spherical symmetry.

Deep in the Vainshtein radius, the force again saturates at the constant value $F_\phi = - m \beta \Lambda^2 / M_P$. This yields the scalar to gravitational force ratio of
\begin{align}
  \frac{F_\phi}{F_G} = 2 \beta^2 \frac{r^2}{r_v^2} \,.
\end{align}
This is very similar to the form of the Galileon force.

The screening curves for this model are plotted alongside the Galileon results in Fig. \ref{fig:plot}.

\section{Discussion}
\label{sec:discussion}

In this work we have derived the flat space solutions for theories with Vainshtein screening mechanisms around planar, cylindrical and spherical sources.  We have considered Galileon theories as a typical example of $(\partial\partial \phi)$ screening and D-BIons as an example of $(\partial \phi)$ screening. Whilst the sources considered in this work represent a tiny subset of all the possible shapes that one could imagine for matter sources, they are sufficient to describe what is happening on large cosmological scales, where almost all matter lives either in walls, filaments or halos.

For the Galileon, there is no screening at all around a planar source, making such structures the best place to look for Galileon fields. Both cylindrical and spherical sources possess a Vainshtein radius within which the scalar force is screened. In the cylindrical case, the ratio of the Galileon force to the gravitational force scales as $r/r_v$ well within the Vainshtein radius, whereas the screening for spherical sources is more efficient, with ratios of either $(r/r_v)^{3/2}$ or $(r/r_v)^2$ depending on whether the cubic or quartic Galileon terms are dominant. Thus, Vainshtein screening is less efficient at hiding the scalar force for cylindrical sources than it is for spherical sources.

For a static system the quintic Galileon term never contributes to the equations of motion, and so observations of static systems can never constrain the Galileon parameter $\lambda_5$. We have shown that the quartic Galileon never contributes to the cylindrically symmetric Galileon equation of motion, and it has been previously shown that none of the Galileon operators contribute to the equation of motion for the field around a planar source. Therefore, if it were possible to measure the Galileon field profile around cosmological walls, filaments and halos, it would be possible to break the degeneracies between the Galileon parameters and determine $\beta$, $\Lambda$ and $\lambda_4$. Information about $\lambda_5$ can only be ascertained from four-dimensional dynamics.

In contrast, for a D-BIonic scalar there is always a Vainshtein radius (or more precisely, a Vainshtein distance scale) governing screening in all the geometries considered. As this does not rely on the symmetries of the D-BIonic Lagrangian, we expect this to be general to all theories with $(\partial \phi)$ screening.
The scalar force is always constant and independent of distance, around an infinite planar source. We find that planar objects are always screened or unscreened, depending on whether or not the width of the source is larger or smaller than the corresponding Vainshtein distance scale. This is in contrast to cylindrical or spherical sources, where only observers within the Vainshtein radius of the source see a screened force. Deep inside the Vainshtein radius, we found that the ratio of the scalar to gravitational forces had the same dependence on $r/r_v$ as the cubic Galileon in cylindrical symmetry and the quartic Galileon in spherical symmetry.

It is interesting to note that the scaling of the Vainshtein radius is quite different in the cylindrical and spherical cases, and also differs between the Galileon and D-Bionic theories. These expressions are displayed together in Table \ref{table:vainshtein}. Compared side-by-side like this, we see that the Galileon scales always contain $M_P \Lambda^3 / \beta$, while the D-BIon scales always contain $M_P \Lambda^2 / \beta$. Up to numerical factors, the Vainshtein radius is simply the combination of the appropriate mass or mass density with these combinations.

\def\arraystretch{2.65}
\begin{table}[t]
\begin{tabular}{|@{\quad}c@{\quad}||@{\quad}c@{\quad}|@{\quad}c@{\quad}|}
  \hline
    Source & Galileon & D-BIon \\
  \hline
  \hline
    Plane & & $\displaystyle \frac{M_P \Lambda^2}{\beta \rho_0}$\\
    \hline
    Cylinder & $\displaystyle \sqrt{\frac{\beta \lambda_0}{M_P \Lambda^3}}$& $\displaystyle \frac{\beta \lambda_0}{M_P \Lambda^2}$\\
    \hline
    Sphere & $\displaystyle \left( \frac{\beta M }{M_P \Lambda^3}\right)^{1/3}$ & $\displaystyle \sqrt{\frac{\beta M}{ M_P \Lambda^2}}$ \\
  \hline
\end{tabular}
\caption{The Vainshtein distance scales in the different theories and symmetries considered in this article. Numerical coefficients have been suppressed in order to demonstrate how the radii scale with various quantities.}
\label{table:vainshtein}
\end{table}
\def\arraystretch{2.4}

\begin{table}[t]
\begin{tabular}{|c||c|c|}
  \hline
    Source & Sphere ($M_\odot$) (pc) & Cylinder ($\lambda_0$) (Mpc) \\
  \hline
  \hline
    Galileon & $\displaystyle 500 \, \beta^{1/3} \left(\frac{10^{-13} \mbox{ eV}}{\Lambda}\right)$ & $\displaystyle \sqrt{\beta} \left(\frac{10^{-13}\mbox{ eV}}{\Lambda}\right)^{3/2}$ \\
    \hline
    D-BIon & $\displaystyle \sqrt{\beta} \left( \frac{10^{-5} \mbox{ eV}}{\Lambda} \right)$ & $\displaystyle \beta \left(\frac{10^{-5}\mbox{ eV}}{\Lambda}\right)^2$ \\
  \hline
\end{tabular}
\caption{Approximate Vainshtein radii for a solar mass sphere and a filament with $\lambda_0 \sim 10^8 M_\odot / \mathrm{Mpc}$ in the Galileon and D-BIon models. For both models we expect $\beta \sim 1$ if the scalar arises from a modification of the gravitational sector.}
\label{table:vainshtein2}
\end{table}

From Fig. \ref{fig:plot}, we see that the D-BIon is somewhat better at screening than the Galileon. However, this statement should be treated cautiously; the plot is shown in units of $r/r_v$, and comparing the Vainshtein radii of different models is a dubious proposition at best. The other thing to note from this figure is that spherical screening is stronger than cylindrical screening within the Vainshtein radius in all cases, which suggests that cylindrical systems may be useful environments in which to search for extra forces.

\subsection{Cosmological implications}

The large scale structure of the universe, sometimes referred to as the cosmic web, is built out of walls, filaments and halos. These are predominantly composed of dark matter, traced by visible galaxies. While the Vainshtein radius of spherical structures like the sun and the galaxy are typically expected to be extremely large, the cylindrical Vainshtein radius may, depending on the parameters, be somewhat reduced compared to spherical expectations. Simulations suggest the existence of filaments of radii $\sim 10 \, \mbox{kpc}$ with nearly constant linear mass densities $\lambda \sim 10^8 \, M_\odot / \mbox{kpc}$ \cite{Harford:2008bw}. Such filaments can be particularly long, with observations suggesting lengths of up to $\sim 100 \, \mbox{Mpc}$ \cite{Bharadwaj:2003xm, Pandey:2010yj}.

We estimate the Vainshtein radii for Galileons and D-BIons around solar mass objects and filaments with the above linear mass density mass in Table \ref{table:vainshtein2}. The reference scale for Galileons $\Lambda = (H_0^2M_P)^{1/3} \sim 10^{-13} \mbox{ eV}$ is chosen because this scale allows the Galileon to be cosmologically relevant at the current epoch \cite{Nicolis2008}, while for D-BIons the scale is taken to be the value which allows the D-BIon to evade lunar laser ranging searches for fifth forces \cite{Burrage:2014uwa}.

We see that for appropriate values of $\beta$ and $\Lambda$, the screening radius for a filament may well be within its radius, although we would typically expect filaments to be screened. The filament screening radii are approximately the same in both models (for the given parameters), at around 100 times the filament radius. This is a significantly smaller ratio than the radius of the sun to its screening radius, which for the D-BIon is around $5 \times 10^7$.

The dependence of Vainshtein screening on the morphology of structures in N-body simulations of the cosmic web has been studied by Falck \textit{et al.} in \cite{Falck:2014jwa}. It was found that dark matter particles in filaments and voids experienced a Galileon force that was unscreened whilst dark matter particles in halos felt a Galileon forces that was screened, compared to the gravitational force they experienced. This supports the analytic results derived here and demonstrates that it is possible to separate cosmological observables by the morphology of the associated cosmological structure.

We have seen that Vainshtein screening is less efficient around objects that are not spherically symmetric. Therefore, the vicinity of walls and filaments may be ideal environments in which to look for the existence of Vainshtein screened fifth forces. If it is possible to observe the motion of particles towards cosmological structures with differing shapes, we may be able to determine whether a fifth force must be screened, and to what degree, around walls, filaments and halos separately. This will allow us to differentiate between $(\partial \phi)$ and $(\partial\partial \phi)$ screening, as the latter is unable to screen walls. It will also allow us to break the degeneracies between the parameters within one class of screening mechanism, as in Galileon models, only the cubic coupling is important around cylindrical sources, while a combination of both the cubic and quartic couplings are important around spherical sources.

\acknowledgments

We thank the Lorentz Center at Leiden University for their gracious hospitality while this work was performed. C.B. is supported by a Royal Society University Research Fellowship. ACD is supported in part by STFC.

\bibliographystyle{utphys}
\bibliography{bibrefs}

\end{document}